\documentclass[aps,prl,preprintnumbers,showpacs,amsmath,amssymb,showkeys]{revtex4-1}
\usepackage[colorlinks=true,linkcolor=blue,citecolor=blue,urlcolor=blue]{hyperref}
\usepackage[utf8]{inputenc}
\usepackage{graphicx}
\usepackage{verbatim}
\usepackage{bm,footnote}

\begin{document}


\title{Phase locked solutions and their stability in the presence of propagation delays }
\author{Gautam C Sethia}
\email{gautam@ipr.res.in}
\author{Abhijit Sen}
\email{abhijit@ipr.res.in}
\affiliation{Institute for Plasma Research, Bhat, Gandhinagar 382 428, India}
\author{Fatihcan M. Atay}
\email{atay@member.ams.org}   
\affiliation{Max Planck Institute for Mathematics in the Sciences, Inselstr.~22-26, Leipzig D-04103, Germany}

\begin{abstract}
We investigate phase-locked solutions of a continuum field of nonlocally
coupled identical phase oscillators with distance-dependent propagation delays. 
Equilibrium relations for both synchronous and traveling wave solutions in the parameter
space characterizing the non-locality and time delay are delineated. For the synchronous states 
a comprehensive stability diagram is presented that provides a heuristic synchronization condition 
as well as an analytic relation for the marginal stability curve. The relation yields simple
stability  expressions in the limiting cases of local and global coupling of the phase oscillators.
\end{abstract}


 
\maketitle


\section{Introduction}
An ensemble of coupled phase oscillators provides a simple yet powerful paradigm 
for studying the collective behaviour of many complex real life systems and has 
been extensively studied in this context\cite{winfree67,ermentrout84,Kuramoto}. 
Under appropriate conditions, phase oscillators
can exhibit stable in-phase synchronous or other type of phase-locked
solutions \cite{ermentrout92}. In a class of discrete networks of oscillators
it has been shown that in-phase synchronous behavior can be achieved even in
the presence of a fixed signal transmission delay \cite{earl03}. More
intricate dynamics have been discovered in nonlocally coupled continuous
networks, whereby phase-locked and incoherent activity can simultaneously
exist at different spatial locations \cite{kuramoto02}, giving rise to a
spatio-temporal pattern that has been termed as a \emph{chimera state }
\cite{abrams04}. In this paper, we discuss the existence and stability of phase-locked solutions 
in a continuum of nonlocally coupled identical phase oscillators with
distance-dependent delays. Such delays are a natural consequence of the finite
speed of information propagation in space. In similar contexts,
distance-dependent delays have been considered in continuum models of neural
activity \cite{crook97,atay05}.
In particular, the effects of distributed delays can be very different from
fixed delays \cite{atay03,atay05,sethia08,sethia10}, and the analysis is generally more
difficult. \newline 
\section{Model system and its synchronous states}
We consider a continuum of identical phase oscillators, arranged on a circular ring $C$
and labeled by $x\in\lbrack-L,L]$ with periodic boundary conditions, whose
dynamics is governed by
\begin{equation}
\frac{\partial}{\partial t}\phi(x,t)  =\omega-K\int_{-L}^{L
}G(z)\sin\left[  \phi(x,t)-\phi\left( x-z,t-\frac{|z|}{v}\right) +\alpha \right]dz, 
\label{pha}
\end{equation}
where $\phi(x,t) \in [0,2\pi)$ is the phase of the oscillator at location $x$ and time $t$, whose intrinsic oscillation frequency is $\omega>0$, $K$ is the coupling strength and $G:[-L,L]\rightarrow\mathbb{R}$ is an even function
describing the coupling kernel.  The quantity $v$ denotes the signal
propagation speed which gives rise to a time delay of $|z|/v$ for distance $|z|$ from  the locations $x$. As the oscillators are located on a ring with circumference $2L$, the distance between any two oscillators is given by the shorter of the Euclidean distance between them along the ring. In this configuration, the maximum distance between the coupled oscillators is $L$ and thus the maximum time delay would be $\tau_m=L/v$. If we denote the location of any two oscillators as $x$ and $x'$ then $z=x-x'$.
$\alpha$ is a parameter which makes the coupling phase-shifted and has been crucial for observing chimera states.

We choose the coupling kernel $G(z)$ to have an exponentially decaying nature and its normalized form is taken to be 
\begin{equation}
G(z)=\dfrac{\sigma}{2(1.0-e^{-L \sigma})}e^{-\sigma|z|}, \label{ker} 
\end{equation}  
where $\sigma > 0$ is the inverse of the interaction scale length and is a measure of the nonlocality of the coupling. The exponential form of $G(z)$  and the $sin$ coupling function are the natural consequences of the reduction of a more general reaction-diffusion system to a phase model under nonlocal and weak coupling limits \cite{kuramoto02}.
We further make time and space dimensionless in  Eqs.(\ref{pha}) and (\ref{ker}) by the transformations $t\rightarrow Kt$, $\omega\rightarrow\omega/K$,
$\kappa \rightarrow \sigma L$, $z \rightarrow z / L$, $\tau_m \rightarrow K\tau_m $ and $x \rightarrow x / L$   and obtain
\begin{equation}
\frac{\partial}{\partial t}\phi(x,t)  =\omega-\int_{-1}^{1
}G(z)\sin\left[  \phi(x,t)-\phi\left( x-z,t-|z|\tau_{m}\right) + \alpha \right]dz. 
\label{phase}
\end{equation}
\begin{equation}
G(z)=\dfrac{\kappa}{2(1.0-e^{-\kappa})}e^{-\kappa|z|} \label{kernel} 
\end{equation}  

We look for phase-locked solutions of Eq.~(\ref{phase}) that have the form:
\begin{equation}
\phi_{\Omega,k}(x,t)=\Omega t+\pi kx+\phi_{0}.\label{sol}%
\end{equation}

These solutions are \emph{phase-locked}, in the sense that the difference of
the phases at two fixed locations in space does not change in time
\cite{ermentrout92}. They can describe spatially uniform equilibria
($\Omega=k=0)$, spatial patterns ($\Omega=0,$ $k\neq0$), synchronous
oscillations ($\Omega\neq0$, $k=0$), or traveling waves ($\Omega\neq0$,
$k\neq0$). Notice that $k\ $needs to be an integer because of periodic
boundary conditions, and the value of $\phi_{0}$ can be taken to be zero by a translation. 

It is useful to first look at the undelayed case ($v=\infty$). Substitution of
(\ref{sol}) into (\ref{phase}) gives the relation between the temporal
frequency $\Omega$ and the wave number $k$ as%
\begin{align}
\Omega &  =\omega-\int_{-1}^{1}G(z)\sin\left(  \pi k z+\alpha\right)  \,dz\nonumber\\
&  =\omega-\hat{G}(k)\sin\alpha\label{nodelay-Omega2}%
\end{align}
where $\hat{G}(k)=$ $\int_{-1}^{1}G(z)\cos(\pi kz)\,dz$ denotes the discrete
cosine transform of $G$. Note that the role of $\omega$ is simply to shift the
value of $\Omega$, so it can be taken to be zero without loss of generality
(we will see later that this is no longer true in the presence of delays).
When $\alpha=0$, $\Omega$ is identical to the individual oscillator frequency
$\omega$. More generally, for $\alpha\in\mathbb{R}$, (\ref{nodelay-Omega2})
yields a unique $\Omega=\Omega(k)$ for any given $k\in\mathbb{Z}$. Hence there
exists a countably infinite set of solutions $\phi_{\Omega(k),k}$,
$k\in\mathbb{Z}$, of (\ref{phase}). The condition for spatial patterns with
wave number $k$ is%
\begin{equation}
\omega=\hat{G}(k)\sin\alpha,\label{pattern-cond}%
\end{equation}
and can only be satisfied if $|\sin\alpha|$ is not too small. Generically,
however, the value of $\Omega$ from (\ref{nodelay-Omega2}) is nonzero, so the
set $\{\phi_{\Omega(k),k}:k\in\mathbb{Z}\}$ includes a unique synchronous
solution ($k=0$) and the rest correspond to traveling waves ($k\neq0$), with
wave speed equal to $\Omega(k)/k$. It turns out that in general only a few of
the solutions $\phi_{\Omega,k}$ can be stable. The linear stability is
determined by the variational equation
\begin{align}
\frac{\partial}{\partial t}u(x,t) &  =-\int_{-1}^{1}G(z) \cos\left( \pi k z+\alpha\right)  \,[u(x,t)-u(x-z,t)]dz\nonumber
\end{align}
where $u(x,t)=\phi(x,t)-\phi_{\Omega,k}(x,t)$. With the ansatz
$u(x,t)=e^{\lambda t}e^{i\pi n x}$, $\lambda\in\mathbb{C}$, $n\in\mathbb{Z}$, we
have%
\begin{equation}
\lambda=-\int_{-1}^{1}G(z)\cos(\pi kz+\alpha)(1-e^{-i\pi n z}%
)dz.\label{nodelay-lambda}%
\end{equation}
After some simplification and using the fact that $G$ is an even function, one
obtains%
\[
\operatorname{Re}\lambda=-(\cos\alpha)\left(  \hat{G}(k)-\frac
{\hat{G}(k+n)+\hat{G}(k-n)}{2}\right)  .
\]
Note that the pair $(\lambda,n)=(0,0)$ is always a solution, corresponding to
the rotational symmetry of the solutions (\ref{sol}). Hence, $\phi_{\Omega,k}$
is linearly asymptotically stable if and only if the right hand side of
(\ref{nodelay-lambda}) is negative for all nonzero integers $n$. For instance,
if $\hat{G}$ has a global maximum (respectively, minimum) at $k$, then the
corresponding solution $\phi_{\Omega,k}$ is stable provided $\cos\alpha>0$
(resp., $<0$). Furthermore, if $\hat{G}(k)>4|\hat{G}(n)|$ for all $n\neq k$,
then $\phi_{\Omega,k}$ is the only stable phase-locked solution (up to a phase
shift) for $\cos\alpha>0$. In particular, if the coupling kernel is constant
(that is, $\hat{G}(k)>0$ iff $k=0$) or has a predominant constant component
($\hat{G}(0)>$ $4|\hat{G}(n)|$, $\forall$ $n\neq0$), then the synchronous
solution $\phi_{\Omega(0),0}$ is the only stable phase-locked solution (up to
a phase shift) for $\cos\alpha>0$, which explains the widespread use of
globally coupled oscillators in synchronization studies.

We now show that the inclusion of propagation delays offers a much richer
solution structure. We again seek solutions $\phi_{\Omega,k}$ of the form
(\ref{sol}) when the propagation velocity $v$ is finite. The relation
(\ref{nodelay-Omega2}) between the temporal frequency and the wave number now
takes the form{{
\begin{equation}
\Omega=\omega-\int_{-1}^{1}G(z)\sin\left[  \Omega \tau_{m} |z|+\pi kz+\alpha
\right]  \,dz,\label{delay-Omega}%
\end{equation}
}}which is an implicit equation in $\Omega$. Note that the right hand side is
a bounded and continuous function of $\Omega$; therefore, (\ref{delay-Omega})
has a solution $\Omega$ for each $k\in\mathbb{Z}$. However, in contrast to the
undelayed case, it is now possible to have several solutions $\Omega$ for a
given wave number $k.$ (Note, however, that the condition for spatial patterns
(\ref{pattern-cond}) remains identical to the undelayed case.) Furthermore,
the linear stability of the phase-locked solutions $\phi_{\Omega,k}$ is
determined through a dispersion relation
\begin{align}
\lambda &  =-\int_{-1}^{1}G(z)\cos\left(  \Omega \tau_{m} |z|+\pi kz+\alpha\right)
\left(  1-e^{-\lambda|z| \tau_{m}}e^{-i\pi n z}\right)  \,dz.\label{delay-lambda}%
\end{align}
As before, linear stability requires $\operatorname{Re}(\lambda)<0$ for all
solutions $\lambda$ of (\ref{delay-lambda}) for all nonzero integers $n$. The
main difference with the undelayed case (\ref{nodelay-lambda}) is that
(\ref{delay-lambda}) is an implicit equation in $\lambda$, so its solution is
not straightforward. Furthermore, even for a fixed value of $n\in\mathbb{Z}$,
(\ref{delay-lambda}) generally has an infinite number of solutions for
$\lambda$.
For simplicity we take $\alpha=0$ and carry out the integration in equation (\ref{delay-Omega}) for $k=0$ to get an analytical equation for the synchronous solutions  $\Omega$ as:  
\begin{align}
\Omega &=\omega-\int_{-1}^{1}G(z)\sin\left(  \Omega\tau_{m}|z|
\right)   \,dz \nonumber\\
 &=\omega-\frac{\kappa}{(1-e^{-\kappa})}\left[ \frac{\Omega \tau_{m}-\Omega \tau_{m}e^{-\kappa}\cos(\Omega \tau_{m})-\kappa e^{-\kappa} \sin(\Omega \tau_{m})}{\kappa^{2}+(\Omega\tau_{m})^{2}}\right], \label{eqlbm}
\end{align}
which is an implicit equation in $\Omega$. Being a transcendental equation its solution can in principle be multi-valued in $\Omega$ for a given set of parameters $\omega, \tau_{m}$ and $\kappa$  and can lead to higher branches of collective frequencies as pointed out by Schuster and Wagner \cite{schuster89} for a system of two coupled oscillators. 
A similar (much lengthier) analytical expression can also be obtained for finite values of $k$ which we do not write here.
We further define a mean delay parameter by
\begin{equation}
\bar{\tau}=\int_{-1}^{1}G(z)\tau_{m}|z|\, dz\
\label{mtaueqn}
\end{equation}
which weights the individual delays with the corresponding connection
weights. With the exponential connectivity given by Eq.~(\ref{kernel}), this translates
into
\begin{equation}
\bar{\tau}=\tau_{m}\frac{e^{\kappa}-\kappa-1}{\kappa(e^{\kappa}-1)}.
\label{mtau}
\end{equation}
This gives values for the limiting cases: $\bar{\tau}\to0$
for local ($\kappa\to \infty$) and $\bar{\tau}\to\tau_{m}/2$ for global ($\kappa\to 0$) coupling.


Fig.~1(a) and Fig~1(b) show plots of the numerical solutions of the equilibrium relations for $\Omega$ as a function of $\bar{\tau}$ for traveling wave with $k=1$and for synchronous ($k=0$) states respectively. In both cases the value of $\kappa$ is taken to be $2.0$ and the plots are obtained for several values of $\omega$. 
In these figures as the curves for $\omega=0.8$ show, it is possible to have multiple solutions $\Omega$ for a given value of $\bar{\tau}$. The stability of these higher states will be discussed in later sections of the paper. One also notes that the lowest branch shows frequency suppression as a function of the mean time delay $\bar{\tau}$.

\begin{figure}
\begin{tabular}
[c]{cc}
\includegraphics[width=0.5\textwidth]{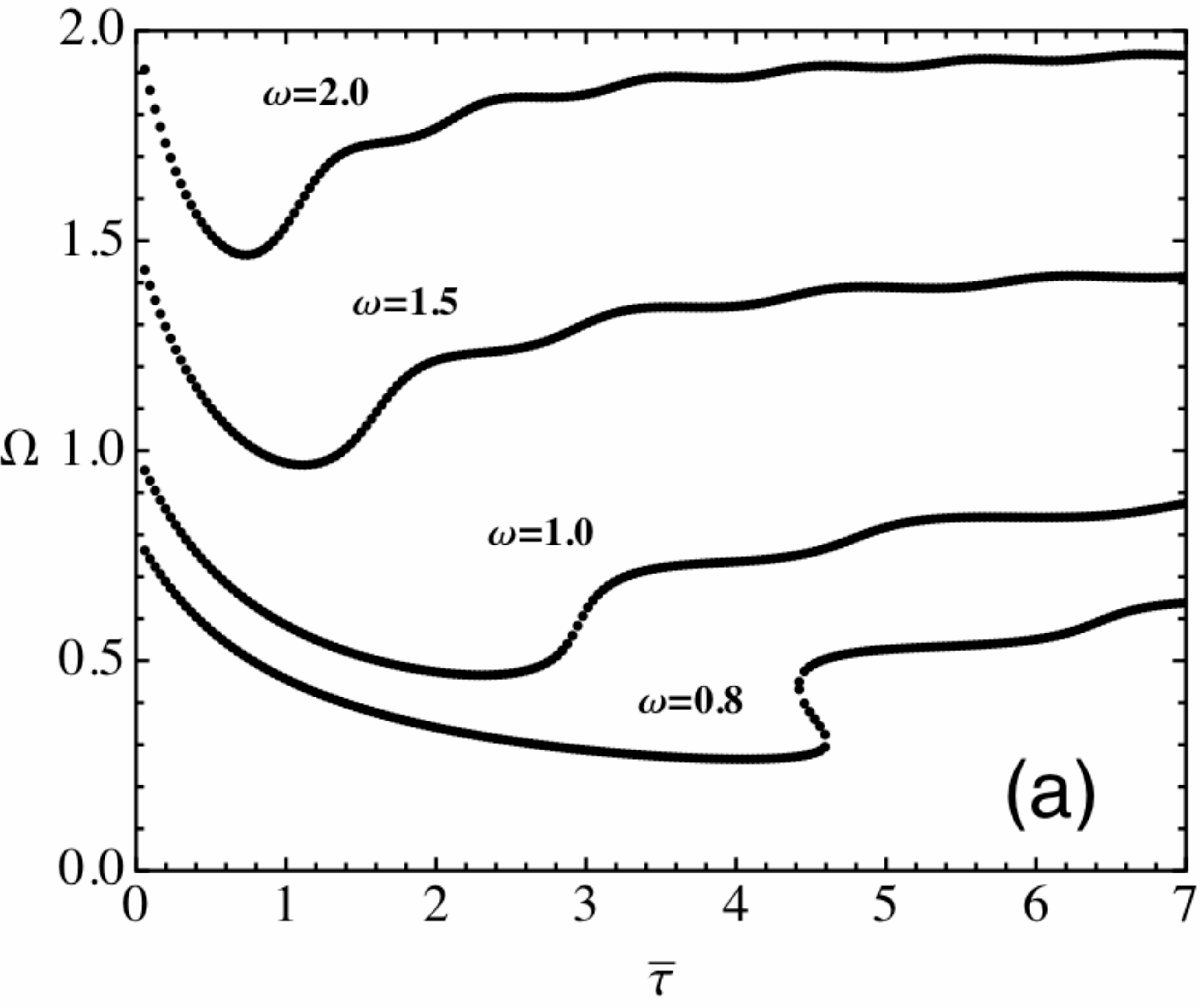}
&
\includegraphics[width=0.5\textwidth]{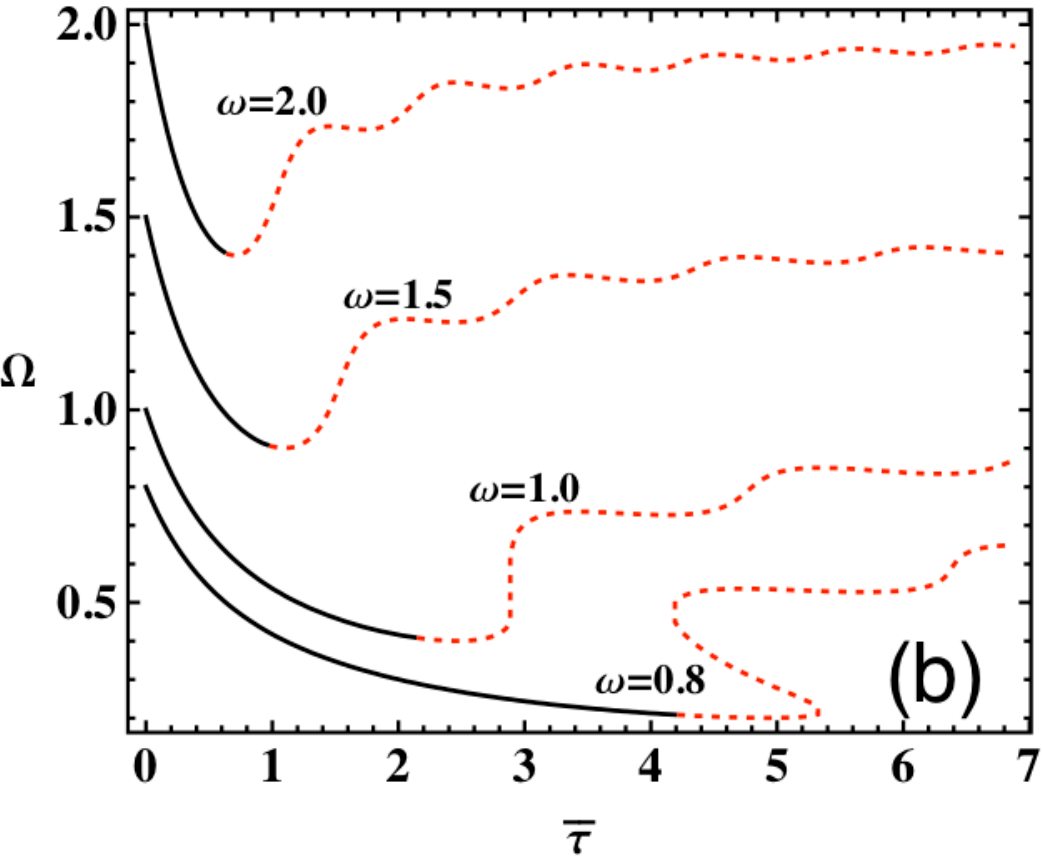}

\end{tabular}
\caption{Frequency $\Omega$ of the phase-locked solutions given by  of Eq.~(\ref{delay-Omega}) is plotted as a function of the mean time delay $\bar{\tau}$ for $\kappa=2.0$ and $\alpha=0.0$. The panel (a) on the left is for  the travelling wave number $k=1$ whereas the panel (b) on right is for synchronous solutions with $k=0$. The different curves correspond to  different values of the intrinsic oscillator frequency $\omega$. The solid portions denote stable states and the dotted ones (in red) unstable states in panel (b).}
\label{fig:fig1}
\end{figure}

\section{Stability of the synchronous solutions}

We now examine the stability of the synchronous solutions \cite{sethia10}, $\phi_{\Omega,0}$ by using the 
eigenvalue equation (\ref{delay-lambda}) obtained in the previous section. 
Writing $\lambda = \lambda_R + i\lambda_I$ and separating Eq.~(\ref{delay-lambda}) into its real and imaginary parts we get
\begin{align}
\lambda_R & = -\int_{-1}^{1}G(z)\cos\left(  \Omega |z|\tau_{m}\right)\left[  1-e^{-\lambda_R|z|\tau_{m}}\cos\left( \lambda_I |z|\tau_{m} + \pi nz\right)\right]dz,\label{lambda_R}\\
\lambda_I & = -\int_{-1}^{1}G(z)\cos\left(  \Omega |z|\tau_{m}\right)e^{-\lambda_R|z|\tau_{m}}\sin\left(\lambda_I|z|\tau_{m} + \pi nz\right)dz.\label{lambda_I}
\end{align}
Since the perturbations $u(x,t)$ corresponding to $n=0$ again yield synchronous solutions, 
the linear stability of the synchronous state requires that all solutions of Eq.~(\ref{delay-lambda}) have $\lambda_R < 0$ for all non-zero integer values of $n$. The marginal stability curve in the parameter space of $(\kappa,\tau_m)$ is defined by $\lambda_R=0$ and in principle can be obtained by setting $\lambda_R=0$ in Eq.~(\ref{lambda_R}), solving it for $\lambda_I$ and 
substituting it in Eq.~(\ref{lambda_I}). In practice it is not possible to carry out such a procedure analytically for the integral Eqs.~$(\ref{lambda_R},\ref{lambda_I})$ and one needs to adopt a numerical approach, which is discussed in  the next section. 


To systematically determine the eigenvalues of Eq.~(\ref{delay-lambda}) in a given region of the complex plane we use multiple methods in a complementary fashion. 
First, Eq.~(\ref{eqlbm}) is solved for
$\Omega$ for a given set of values of the parameters $\kappa$, $\omega$ and $\tau_{m}$. Following that, we need to determine the complex zeros of the function $f(\lambda)$ defined as
$$ f(\lambda) = \lambda + \int_{-1}^{1}G(z)\cos\left(  \Omega\tau_{m}|z|\right) 
\left(  1-e^{-\lambda|z|\tau_{m}}e^{-i\pi n z}\right)  \,dz,$$
which is equivalent to finding solutions $\lambda$ of (\ref{delay-lambda}).
To do this we have primarily relied on the numerical technique developed by Delves and Lyness \cite{delves67} based on the Cauchy's argument principle. By this principle the number of unstable roots $m$ of $f(\lambda)$ is given by
$$ m = \frac{1}{2\pi i}\oint_C\frac{f'(\lambda)}{f(\lambda)}d\lambda, $$
where the closed contour $C$ encloses a domain in the right half of the complex $\lambda$ plane with 
the imaginary axis forming its left boundary. Once we get a finite number for $m$ we further trace the location of the roots by plotting  the zero value contour lines of the real and imaginary parts of the function $f(\lambda)$ in a finite region of the complex plane $(\lambda_R,\lambda_I)$. 
The intersections of the two sets of contours locate all the eigenvalues of Eq.~(\ref{delay-lambda}) in the given region of the complex plane. The computations are done on a fine enough grid (typically $80 \times 80$) to get a good resolution. A systematic scan for unstable roots is made by repeating the above procedure for many values of the perturbation number $n$ and by gradually extending the region of the complex plane. We have made extensive use of {\it Mathematica} in obtaining the numerical results on the stability of the synchronous states.


In Fig.~\ref{fig:fig1} the solid portion of the curve shows the stable synchronous states of 
Eq.~(\ref{phase}) for $\kappa=2.0$ and for various values of $\omega$ and $\bar{\tau}$. The terminal point on a given  solid curve of $\Omega$ {\it vs} $\bar{\tau}$ marks the marginal stability point. The marginal stability point is seen to shift towards larger values of $\bar{\tau}$  as one moves down to curves with lower
values of $\omega$. 
\begin{figure}
\includegraphics[width=0.8\textwidth]{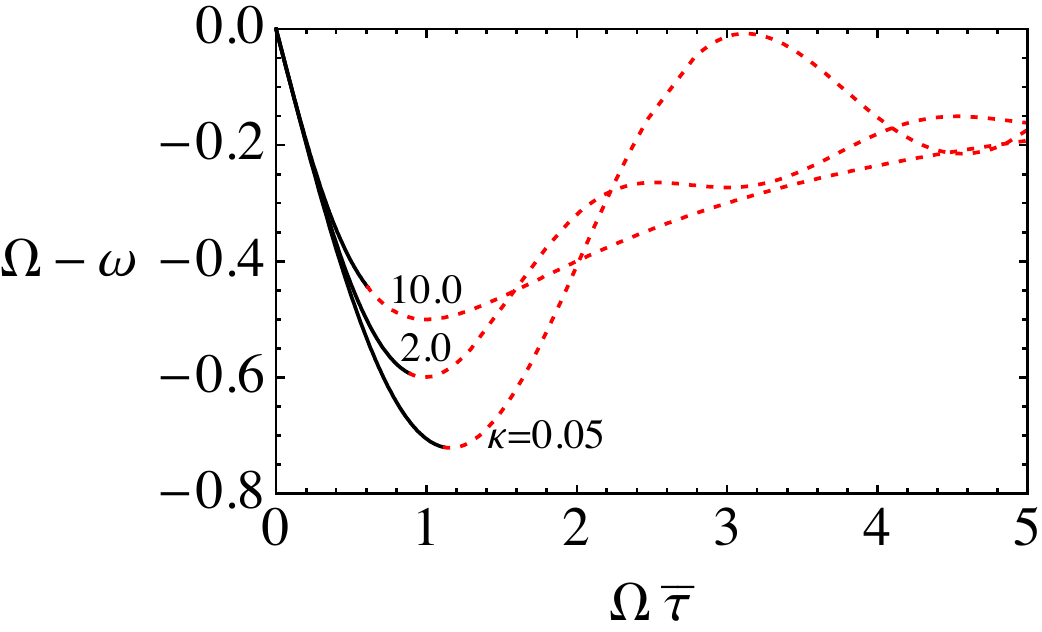}
\vskip.0cm
\caption{Solutions of Eq.~(\ref{eqlbm}) for the synchronous oscillation frequency for several values of $\kappa$, plotted in terms of $\Omega - \omega$ versus $\Omega \bar{\tau}$. Note that $\Omega-\omega=H(\Omega \bar{\tau},\kappa)$ for the equilibrium solutions (see Eq.(\ref{H})). In this representation, the different curves of Fig.1 corresponding to different values of $\omega$  collapse onto a single curve for a given value of $\kappa$. The solid (black) portions of the curves correspond to stable synchronous states and the dotted ones (red) to unstable synchronous states.}
\label{fig:fig2}
\end{figure}
A more compact representation is obtained if one plots 
$\Omega-\omega$ versus $\Omega\bar{\tau}$, since in this case the solutions corresponding to different values of $\omega$ for a given $\kappa$ consolidate onto a single curve,
as shown in Fig.~\ref{fig:fig2} for $\kappa= 0.05,2.0$ and $10.0$ respectively. 

It is seen from both figures that the stability domains of the synchronous solutions are restricted to the lowest branch where the curves are decreasing.
This suggests a heuristic necessary (but not sufficient) condition for stability of the synchronous solutions: From Fig.~\ref{fig:fig1} we have that 
$\partial\Omega / \partial \bar\tau < 0$ for stable synchronous solutions, and
from Eq.~(\ref{eqlbm}) we calculate
\begin{equation}  \label{d-omega}
	\frac{\partial\Omega}{\partial \bar{\tau}} = 
	\frac{-\Omega c_\kappa I} {1 + c_\kappa\bar{\tau} I}
\end{equation}
where
\begin{equation}  \label{I}
	I =  \int_{-1}^{1} |z| G(z) \cos(c_\kappa \Omega \bar\tau |z|) \,dz
\end{equation}
and
\[
	c_\kappa = \frac{\kappa(e^\kappa - 1)}{e^\kappa - 1 - \kappa}.
\]
Since $c_\kappa > 0$ and $I$ is bounded, the denominator in Eq.~(\ref{d-omega}) is positive for small values of $\bar{\tau}$. Hence, for positive $\Omega$, the requirement $\partial\Omega / \partial \bar\tau < 0$ implies the condition $I>0$, 
that is,
\begin{equation}
\int_{-1}^{1}|z|G(z)\cos\left( c_\kappa \Omega \bar\tau |z|\right)dz\;>\;0
\label{nec_cond}
\end{equation} 

An alternative approach to arrive at the necessary condition (Eq.\ref{nec_cond}) would be to make use of the results presented in Fig.\ref{fig:fig2}. We recast the dispersion relation given by  Eq.(\ref{eqlbm}) in the form:
\begin{equation}
 \Omega-\omega = H(\Omega \bar{\tau},\kappa)
\label{H}
\end{equation} 
where
$$H(\Omega \bar{\tau},\kappa) = -\int_{-1}^{1}G(z)\sin\left( c_\kappa \Omega \bar\tau  |z|\right) dz $$ 
It is seen from Fig.\ref{fig:fig2} that the stability domain of the synchronous solutions is restricted to the lowest branch where the curves have a negative slope. This again suggests a heuristic necessary condition for stability of the synchronous solutions to be $H'<0$ leading to the necessary condition given by Eq.(\ref{nec_cond}). The prime indicates a derivative of $H(\Omega \bar{\tau},\kappa)$ $w.r.t$ $\Omega \bar{\tau}$. 

As we will see later, the marginal stability curve obtained from $H'$ or $ I = 0$ does lie above the true marginal stability curve (see Fig.~\ref{fig:fig4}), confirming that
condition (\ref{nec_cond}) is necessary but not sufficient for the stability of synchronous states.
\begin{figure}
\begin{tabular}[c]{cc}
\includegraphics[width=0.8\textwidth]{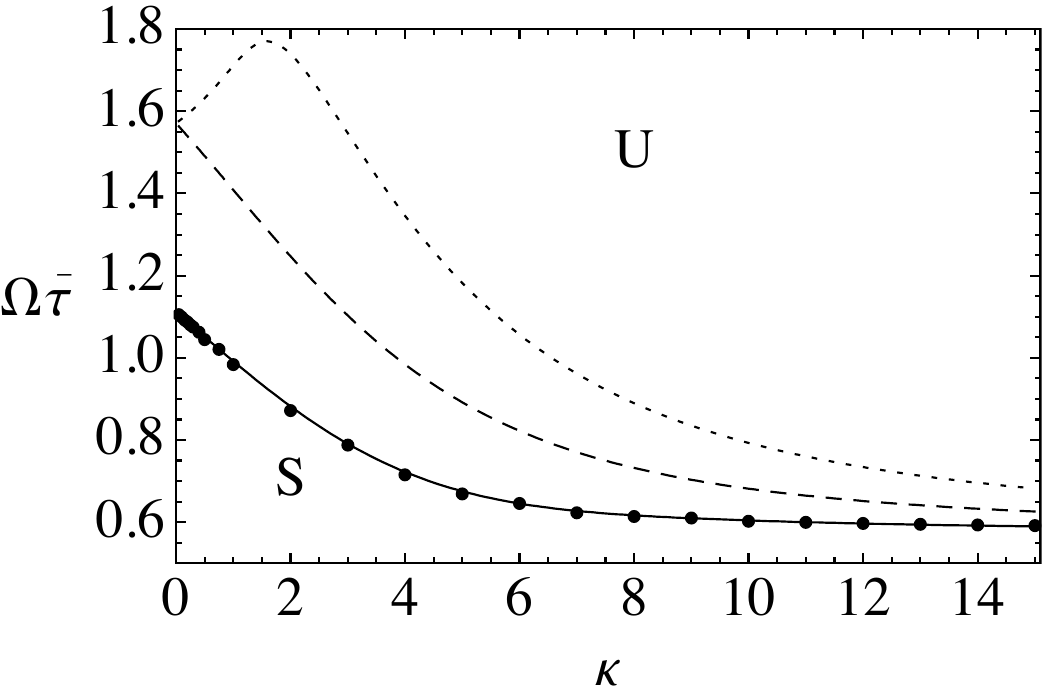}
\end{tabular}
\caption{The marginal stability curve (solid curve) in the ($\Omega\bar{\tau},\kappa$) space, obtained from the lowest branch solutions of Eq.~(\ref{marg_stab}) for $n=1$. The filled
circles correspond to numerical results from eigenvalue analysis of Eq.~(\ref{delay-lambda}), and show a perfect fit to the analytical result. The dashed and dotted curves correspond to marginal stability curves
obtained for $n=2$ and $n=3$ perturbations, respectively. The symbols $S$ and $U$ denote stable and unstable regions in the parameter space.}
\label{fig:fig3}
\end{figure}
\begin{figure}
\begin{tabular}[c]{cc}
\includegraphics[width=0.8\textwidth]{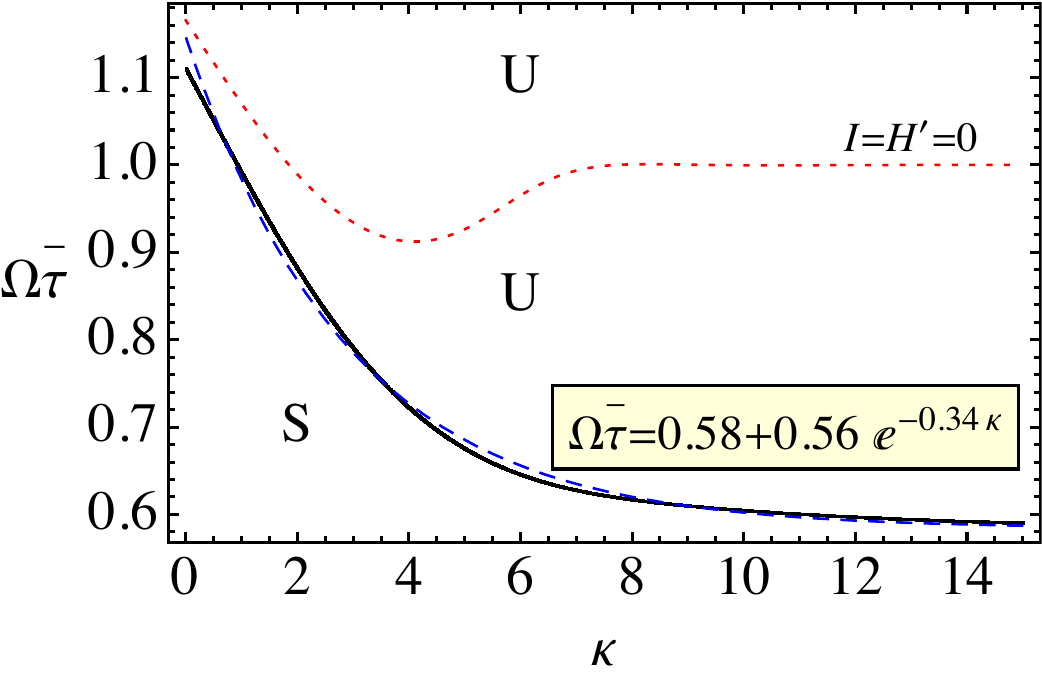}
\end{tabular}
\caption{ A numerical fit to the marginal stability curve gives an approximate scaling law in the form of an offset exponential relation between $ \Omega \bar{\tau}$ and $\kappa$. The marginal stability curve (solid, in black) of Fig.~{\ref{fig:fig3}} has been replotted along with the fitted curve (dashed, in blue).  The dotted curve (in red) is obtained from the condition $H'$ or $I=0$.}
\label{fig:fig4}
\end{figure}

The points where solid and dotted lines meet in the curves of Fig.~\ref{fig:fig2} mark the marginal stability point for the respective $\kappa$ values. These  points are obtained for a range of $\kappa$ values and are plotted in ($\Omega \bar{\tau},\kappa $) space in Fig.~\ref{fig:fig3} by filled points.  They all lie on a single curve, which is analytically derived below.
Our numerical results further reveal that for the marginal stability points, the imaginary part of the eigenvalue of the mode is zero---in other words the mode loses stability through a saddle-node bifurcation.  It can easily be checked by inspection that $\lambda_I=0$ is one of the  solutions of Eq.~(\ref{lambda_I}) for any value of $\lambda_R$; 
however, it is not evident analytically that this is the only possible solution for $\lambda_R=0$, and our numerical results have helped us confirm that this is indeed the case.
Hence, putting $\lambda_R =\lambda_I =0$ in Eq.~(\ref{lambda_R}) we get the following integral relation between the parameters $\Omega, \tau_m$ and $\kappa$.
\begin{equation}
\int_{-1}^{1}G(z)\cos\left(  \Omega \tau_{m} |z|\right)\left[  1-\cos\left(\pi n z\right)\right]dz = 0
\label{marg_stab}
\end{equation}
Further, we have also observed that the most unstable perturbation is the one with the lowest mode number, namely $n=1$. Therefore Eq.~(\ref{marg_stab}) with $n=1$ defines the marginal stability curve, so the condition for synchronization takes the form
\begin{equation}  
   \int_{-1}^{1}G(z)\cos\left(  \Omega \tau_{m} |z|\right)\left[  1-\cos\left(\pi z\right)\right]dz > 0
\label{marg_stab2}
\end{equation}
The solid line in Fig.~\ref{fig:fig3} is the analytical curve of marginal stability defined by setting the left side of Eq.~(\ref{marg_stab2}) to zero, and it can be seen that the numerically calculated marginal values (represented by points) fit this curve perfectly. The figure also shows the stability curves obtained for the $n=2$ and $n=3$ perturbations 
(dashed and dotted lines, respectively) and these are seen to lie above the $n=1$ marginal stability
curve. We have carried out a numerical check for a whole range of higher $n$ numbers and the results are consistent with the above findings. 

For a system with constant delay $\tau$ (i.e. if $\tau_m|z|$ is replaced by $\tau$ in Eq.~(\ref{phase})), the $\cos(\Omega\tau)$ term can be taken outside of the integral in (\ref{marg_stab2}) and the remaining integrand is nonnegative. Hence, the synchronization condition in this case becomes simply 
\begin{equation}  \label{strogatz-cond}
 \cos(\Omega\tau)>0.	
\end{equation}
This agrees with the results obtained previously  for constant-delay systems  \cite{yeung99,earl03}.Thus our result, as given by Eq.~(\ref{marg_stab}), generalizes the condition (\ref{strogatz-cond}) to systems with space-dependent delays, and shows  a nontrivial relation between the spatial connectivity and delays for the latter case. 


In order to gain some intuition into the complex interaction between connectivity and delays, we have obtained an approximate expression for the marginal stability curve by a numerical fitting procedure, yielding the relation 
\begin{equation}  \label{fit}
	\Omega \bar{\tau} < 0.58 +0.56 e^{-0.34 \kappa}
\end{equation}
for the stability of synchronous oscillations.
Here, the left side involves the temporal scales of the dynamics (namely, it is the average time delay normalized by the oscillation period of the synchronized solution) while the right hand side involves the spatial scales of connectivity. In this view, the synchronization condition is a balance between the temporal and spatial scales.
For high connectivity ($\kappa\to 0$), the system can tolerate higher average delays in maintaining synchrony, and the largest allowable delays decrease roughly exponentially as the spatial connectivity is decreased. In the same figure we have also plotted with dotted curve the approximate condition (\ref{nec_cond}), which is found to lie above  the marginal stability curve in the entire range of $\kappa$. The disparity between the two curves becomes 
particularly noticeable  at large values of $\kappa$.
\section{Stability conditions in the limiting cases}
Since the marginal stability curve given by the condition Eq.~(\ref{marg_stab2}) is an analytical one, one can obtain the limiting values of the phase shifts ($\Omega \bar{\tau}$ ) in the global ($\kappa\to 0$) as well as in the local limits ($\kappa\to \infty$). They are given as: 
\begin{equation}
\Omega \bar{\tau}<\frac{\pi}{2\sqrt{2}} =1.11072 
\end{equation}
in the case of global coupling and 
\begin{equation}
\Omega \bar{\tau}<\frac{1}{\sqrt{3}}= 0.57735
\end{equation}
in the case of local coupling.

Similarly one can obtain the limits on the frequency depression for the stability of the corresponding synchronous state in the two limiting cases (see Fig.~\ref{fig:fig2}). In the case of global coupling the condition becomes:
\begin{equation}
\Omega-\omega > -\frac{\sin ^{2}(\frac{\pi}{2\sqrt{2}})}{\frac{\pi}{2\sqrt{2}}} = -0.722819
\end{equation}
whereas in the case of local coupling the condition is :
\begin{equation}
\Omega-\omega > -\dfrac{\sqrt{3}}{4} = -0.4433013
\end{equation}

We note from Fig.~\ref{fig:fig2} that the disparity between $\Omega$ and $\omega$ ($|\Omega-\omega|$) increases with delay induced phase shifts ($\Omega \bar{\tau}$ ) and at some critical point the synchronous state becomes unstable. The value of the disparity at this critical juncture is larger for higher connectivity (global) and smaller for lesser connectivity (local) as we see from above.

\section{Conclusions and Discussion}
We have investigated the existence and stability of the synchronous solutions of a continuum of nonlocally coupled phase oscillators with distance-dependent time delays. Our model system is a generalization of the original Kuramoto model by the inclusion of naturally occurring propagation delays. The existence regions of the equilibrium synchronous and traveling wave solutions of this system are delineated.  The equilibrium solutions of the lowest branch are seen to exhibit frequency suppression as a function of the mean time delay. We have carried out a linear stability analysis of the synchronous solutions and obtained a comprehensive marginal stability curve in the parameter domain of the system. Our numerical results show that the synchronous states become unstable via a saddle-node bifurcation process and the most unstable perturbation corresponds to an $n=1$ (or kink type) spatial perturbation on the ring of oscillators. These findings allow us to define an analytic relation, given by Eq.~(\ref{marg_stab2}), as a condition for synchronization. We have also obtained approximate forms for the synchronization condition that provides a convenient means of assessing the stability of synchronous states. Our results indicate an intricate relation between synchronization and connectivity in spatially extended systems with time delays. A detailed analysis of the stability of traveling wave solutions is currently in progress and will be reported later.

\bibliographystyle{pramana}



%

\end{document}